# Electronic ferroelectricity in carbon based materials

Natasha Kirova[1,3*] and Serguei Brazovskii[2,3]

[1]*LPS, CNRS, Univ Paris-Sud, Université Paris-Saclay, 91405 Orsay Cedex, France*
[2]*LPTMS, CNRS, Univ Paris-Sud, Université Paris-Saclay, 91405 Orsay Cedex, France*
[3]*Moscow Institute for Steel and Alloys, Leninskii av. 4, 119049 Moscow, Russia.*

We review existing manifestations and prospects for ferroelectricity in electronically and optically active carbon-based materials. The focus point is the proposal for the electronic ferroelectricity in conjugated polymers from the family of substituted polyacetylenes. The attractive feature of synthetic organic ferroelectrics is a very high polarizability coming from redistribution of the electronic density, rather than from conventional displacements of ions. Next fortunate peculiarity is the symmetry determined predictable design of perspective materials. The macroscopic electric polarization follows ultimately from combination of two types of a microscopic symmetry breaking which are ubiquitous to qusi-1D electronic systems. The state supports anomalous quasi-particles - microscopic solitons, carrying non-integer electric charges, which here play the role of nano-scale nucleus of ferroelectric domain walls. Their spectroscopic features in optics can interfere with low-frequency ferroelectric repolarization providing new accesses and applications. In addition to already existing electronic ferroelectricity in organic crystals and donor-acceptor chains, we point to a class of conducting polymers and may be also to nano-ribbons of the graphene where such a state can be found. These proposals may lead to potential applications in modern intensive searches of carbon ferroelectrics.

## 1. Introduction

Ferroelectricity is a phenomenon of spontaneous controllable electric polarizations in some, usually crystalline, solids [1]. Ferroelectric materials have a wide range of applications attracting related fundamental studies. They are used as active gate materials and electric RAM in microelectronics, as capacitors in portable WiFi communicators, electro-optical-acoustic modulators, electro-mechanical actuators and transducers, and ultrasound sensors in medical imaging [2-4]. For the last demand, flexible lightweight carbon ferroelectrics would be particularly welcomed.
Traditional ferroelectrics are inorganic materials, or crystals of inert organic molecules (e.g. Rochelle salt $KNaC_4H_4O_6 \cdot 4H_2O$ or triglycine sulfate $(NH_2CH_2COOH)_3 \cdot H_2SO_4$ )**,** but new organic families started to appear in 2000s [5-7]. Among other virtues like the lightness, organic materials are thought to be free from the processing difficulty of conventional ferroelectrics, like compatibility and contamination [8]. An obstacle to employing the new organic ferroelectrics was that the ferroelectric phase existed only well below the room temperature – until the recent breakthrough [7].
Beyond organic crystals, old and new, there is also a special demand for plastic ferroelectrics, e.g. as sensors for the acoustic imaging, in shapeable actuators and capacitors, etc. Their important applications would be in medical imaging, because of a request for low weight material which acoustic impedance would be compatible with biological tissues. (Until now, this demand was satisfied by exploiting composite materials based on powders of ferroelectric oxides in a polymeric matrix which sensitivity is rather low.) The most exploited material is Poly(vinylidene fluoride) (PVDF) [9]. The advantage of all-polymeric ferroelectrics is that they are light, flexible, non toxic, unique as long stretching actuator. They are cheap to be produced being solvable in common organic solvents, so the active elements can be processed by a simple spin-casting. Still, polarizabilities of existing plastic ferroelectrics are rather low, providing the dielectric permittivity $\varepsilon \sim 10$, in comparison with $\varepsilon \sim 500$ in traditional inorganic ferroelectrics, hence a modest efficiency of devices.

---

[*] Corresponding author, e-mail: kirova@lps.u-psud.fr

The ferroelectric transition in PVDF and alike is driven by a cooperative ordering from cis to trans conformations of the polymeric chain, where chains' rotations give rise to opposite displacements of differently charged ligands – here the hydrogen and the fluorine, so in effect the electric polarization comes from ionic displacements as in inorganic materials.

The known purely polymeric ferroelectrics belong to the class of saturated polymers which do not possess the delocalized π - electrons, so the material is not active either electronically or optically. A synergy of all those properties can take place in π-conjugated synthetic carbon based materials like organic conductors and donor-acceptor chains or conducting polymers. The principle novelty is that in these materials the polarization comes from ordering of electrons rather than from displacements of ions, hence the nickname "electronic ferroelectricity" [10].

Beyond a set of principally new features, the responsive π- electronic systems can greatly enhance the dielectric permittivity ε. This expectation has already been confirmed in the family of organic conducting crystals (TMTTF)$_2$X – the Fabre salts: the background dielectric permittivity is ε ~ $10^3$ with astonishing rise to ε ~ $10^6$ at the phase transition. The high permittivity coexists with a low-activated conductivity while keeping the spin paramagnetism; altogether that gives rise to a new state of the "ferroelectric narrow-gap Mott semiconductor" [5,11-13]. The screening by remnant carriers eliminates the usual hysteresis which leads to fast repolarization. The frequency dependence of ε allows separating the critical relaxation within ferroelectric domains and sweeping of domain walls, to extract the critical slowing-down near the transition temperature and the low frequency absorption coming from the creep of domain walls [14]. Less conducting electronic ferroelectrics have been realized as organic donor-acceptor complexes with neutral-ionic transitions where the ferroelectricity [6,15,16] and its electronic nature [10] have been directly confirmed. Also, these materials became the objects of studies by advanced femto-second and tera-Hz optics [17-21].

The microscopic picture of new ferroelectric materials is based on coexisting symmetry lowering effects. The ground state degeneracy brings to life anomalous elementary excitations - the microscopic solitons, which carry non-integer electric charges and play the role of nucleus, at nano-scales, ferroelectric domain walls [22-26].

In this review, we discuss the carbon-based electronic ferroelectrics: from already established ferroelectricity in organic crystals and bimolecular donor-acceptor chains to its possible existence in conducting polymers and graphene nano-ribbons. We argue that a combination of two types of microscopic symmetry breaking in qusi-1D electronic systems should lead to a state with the macroscopic electric polarization. In addition to existing ferroelectricity in organic crystals [5,11] and donor-acceptor chains (see [10] and rfs. therein), we indicate [27,28] a class of conducting polymers and may be also nano-ribbons of the graphene where such a ferroelectric state can be found.

The paper is organized as follows: In Ch. 2 we present the basic introduction to the ferroelectric state. In Ch. 3 we discuss the symmetry defined way to construct the ferroelectrics. In Chs. 4-6 we consider cases of the electronic ferroelectricity in conducting polymers, organic crystals and bi-molecular chains. In Ch. 7 we describe shortly a theory of the ground state and of elementary ferroelectric domain walls - solitons with noninteger charges. Ch. 8 is devoted to the summary, future road map, and conclusion.

**2. Basic introduction to ferroelectricity.**

When an external electric field *E* is applied to a dielectric, positively/negatively charged species shift or rotate or align along/against the field. The density of induced dipole moment is called the polarization *P* which appears in several forms [1]:

- *electronic polarization* exists in all dielectrics as the electronic cloud deforms and shifts with respect to a nuclei, so an atom or a molecule acquires an induced dipole moment (Fig.1, left)
- *orientational polarization* exists in substances built from asymmetric molecules, like $SiO_2$, $H_2O$ or organic molecules having the preexisting dipole moment. Molecules reorient their dipoles aligning them to the field (Fig.1, center).
- *ionic polarization* exists in all ionic solids - NaCl, MgO where weekly bound positive and negative ions shift in opposite directions (Fig.1, right)

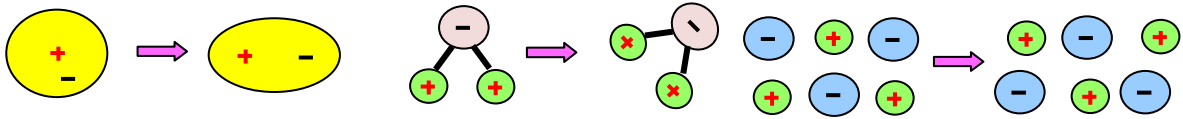

Fig.1. Effects of the external electric field for cases of intra-molecular (left), orientational (center) and ionic displacive (right) polarizations.

At low $E$, the induced dielectric polarization $P$ is proportional to the field (Fig.2, left). But at higher $E$, the polarization growth slows down (see Fig.2); they are called the paraelectrics. In spite of the tendency to saturation, the polarization returns to zero if the electric field is switched off.

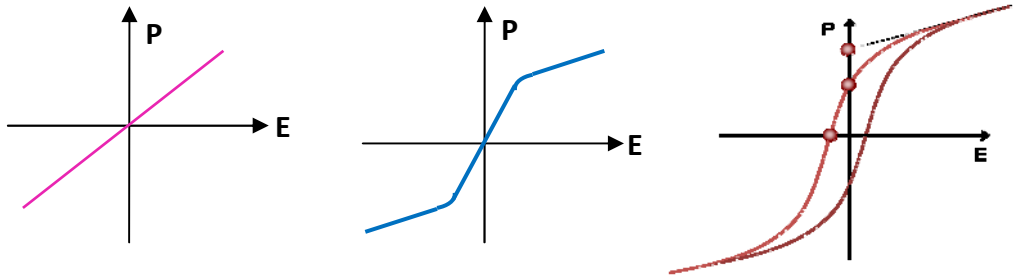

Figure 2. Polarization $P$ versus applied electric field $E$ for a dielectric (left), a paraelectric (center), and a ferroelectric (right).

Some materials are able to generate the macroscopic polarization which happens spontaneously below a transition temperature $T_{FE}$. The local dipoles may be preformed, then the transition is called as the ordering type; otherwise it is called displacive. To be the ferroelectric, the polarization might be able of switching the direction following the applied electric field; if not, it is called the pyroelectric. From the crystallography point of view, a pyroelectric does not possess the inversion center of symmetry at any T, while the ferroelectric looses it only below $T_{FE}$.

Unlike pyroelectrics, in ferroelectrics the polarization direction can be switched by the applied electric field. In real time, the repolarization goes by expanding of favorable domains, hence by propagation of domain walls which can be pinned by lattice defects. Then the polarization depends not only on the present field but also on its history, yielding a hysteresis loop (Fig. 2, right). The value of $P$ persistent at $E=0$ is called the remnant polarization. The value of $P$ extrapolated back from the saturation limit of high $E$ is called the saturated polarization. Regions with different orientations of the polarization vector coexist below the saturation; they are called ferroelectric domains separated by domain walls. The reversed field required for removing the earlier polarization is called the coercive field. Further increase of the reverse field changes the polarization to the opposite, so the hysteresis loop is formed.

There is a common paradox of the ferroelectric state: the polarization works against the Coulomb energy. In traditional inorganic ferroelectrics (PZT - $Pb(Zr_xTi_{1-x})O_3$, SBT - $SrBi_2Ta_2O_9$, etc.) the ferroelectric polarization appears as a result of ions' displacements (Fig.3, left).

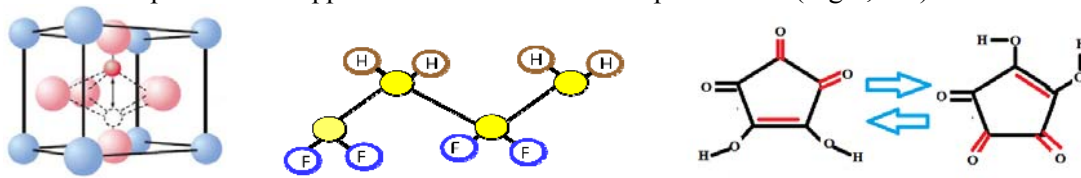

Figure 3. Various types of ionic ferroelectrics. Left: PZT, the arrow indicates a displacement of Ti/Zr ions from the centrosymmetric position. Center: organic polymer PVDF in the trans configuration. Right: molecular conformations giving rise to polarity through the π-bond switching and the intermolecular proton-transfer [7].

In the polymer PVDF (Fig.3, center) the polarization comes from cis-trans isomerization fitting to preferable angles of molecular bonds. In inert organic crystals there are molecular conformations (Fig.

3, right). In all these cases, one cannot make certain predictions about the ferroelectric instability without a thorough microscopic analysis.

Finally, we notice a terminological and historical curiosity: ferroelectric materials do not need to contain the iron or any other atoms with magnetic moments. The prefix "ferro" was given by analogy with ferromagnets which have spontaneous magnetization and exhibit similar hysteresis loops. However, this similarity comes out only at the phenomenological level, and the microscopic mechanisms of a spontaneous electric and magnetic polarization are completely different. The materials possessing both the electric and magnetic polarizations do exist; they form a very important class of multiferroics. This direction is beyond our present consideration; just notice that certain organic ferroelectric conductors fall, at low $T$, to a magnetically ordered spin-density-wave state – in (TMTTF)$_2$X series; see also [29] concerning layered organic compounds. Notice also, that the first ever ferroelectric (discovered in 1920) was an organic crystal – the Rochelle salt. The compound KNaC$_4$H$_4$O$_6$•4H$_2$O have been known since as long as 1672 as the Seignette's salt, named after its discoverer. Curiously, this was a by-product of wine making. In the Russian and former Soviet literature, the name segnetoelectricity is exploited instead of the ferroelectricity.

## 3. Symmetry defined ways to build the electronic ferroelectrics.

Discoveries of electronic ferroelectricity in quasi one-dimensional conducting organic stacks of (TMTTF)$_2$X [5,11] and then in layered BEDT-based compounds [30], and in donor –acceptor chains like TTF-CA [6] have brought to life a new mechanism of the predominantly electronic origin. For electronic ferroelectricity there is a secure way of the design based on symmetry defined effects, driven by electronic correlations like in organic crystals, electron-phonon interactions like in conducting polymers, or both effects together like in donor-acceptor chains TTF-CA.

To achieve the ferroelectric state in a crystal, the inversion symmetry has to be lifted, as well as the mirror symmetry, and the glide plane should be removed. But the double degeneracy of the ground state must be kept; otherwise the state will be a non-switchable pyroelectric. Consider a chain with double dimerization of sites and bonds, see Fig.4. The dimerization of sites means the alternating difference of atoms or molecules or attached ligands (radicals). Alternation of bonds removes the mirror symmetry; alternation of sites removes the glide plane. The local view is that dipoles (indicated by arrows on the Fig. 4) formed by different bonds are not compensated.

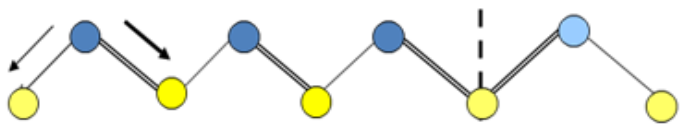

Figure 4. Instruction for the ferroelectric design - combined symmetry breaking. The zig-zag structure is typical but not important (cf. linear Pt-chains compounds).

The microscopic picture of all new materials (see Ch. 7 below) is based upon such coexisting symmetry lowering effects. If both dimerizations are built-in like in alternative stacks of organic crystal (TMTTF)$_{0.5}$(TMTST)$_{0.5}$ [31], then the material is pyroelectric, the polarization is frozen. To achieve the ferroelectricity with its switchable direction of polarization, at least one of these effects should come as a spontaneous symmetry breaking, i.e. any one of dimerizations should be produced by a phase transition. There are the following options:

1. *Dimerization of bonds is built in; dimerization of sites is spontaneous*: it happens in organic conductors (TMTTF)$_2$X with strong electronic correlations. Charge ordering is endorsed by the energy gain from falling of the originally metallic state to the Mott insulator state: band filling changes from 1/4 to 1/2 [5,11,12].
2. *Dimerization of sites is built in; dimerization of bonds is spontaneous:* depending on interactions, we have two possibilities:
a. if electron-lattice interactions dominate, then the Peierls dimerization should take place: that was a proposal for conjugated polymers of the (AB)$_x$ type, expected nowadays to be seen in substituted polyacetylenes (CRCR′)$_x$ [24-26].

b. Another possibility can be given by quasi one-dimensional chains of alternating molecules, each carrying the spin 1/2. Then the spontaneous dimerization of bonds appears as the so called spin-Peierls transition [32].

3. *Both sites' and bonds' dimerizations are spontaneous*: that corresponds to the 1st order neutral-ionic transition in donor-acceptor stacks like TTF-CA [6,10]. Truly, the sites nonequivalence is always present, but it is augmented by the spontaneous intermolecular charge transfer.

## 4. Ferroelectric conjugated polymers

In view of applications, organic crystals are too fragile, and their transition temperatures to the ferroelectric state are still in the cryogenic range (with one exception [7] which does not belong the class of electronic ferroelectrics). So it is tempting to find a similar effect in more robust and technological materials like conjugated polymers.

In general, there are two types of polymers: saturated and conjugated ones. The last are called also conducting, under the great impact of discovery [33] of the metallic polyacetylene. They are optically and electronically active thanks to presence of the π-electrons' manifold. Saturated polymers - traditional plastics like the polyethylene $(CH_2)_x$ are most common in our life. In chemical language these polymers have fully saturated very stable sigma-bonds. In language of physics, all electronic bands are filled, with forbidden gaps 4-5 eV; the polymers are not active either electronically or optically.

The π-conjugated polymers possess, from the chemistry point of view, one carbon bond which is not saturated: the π- bond. In physics language, there is a broad ~8-10 eV band which is half filled by essentially delocalized electrons. The Fermi level is emptified by a relatively small ~1.5-2.5 eV gap which is usually inherent, but in trans-polyacetylene, the $(CH)_x$ chain, the gap is opened as a secondary effect because of the spontaneous bonds' dimerization due to electron-lattice interactions (the Peierls effect, later known as the SSH result [34]). These polymers are 1D semiconductors in a sense that they can be doped, optically pumped, light emitting - even lasing, forming p-n junctions. A broader family of active carbon-based materials includes other π-conjugated systems like fullerenes, nanotubes, and graphene.

## 4.1. Existing ferroelectric inert polymers.

The most studied [9] ferroelectric saturated polymer is $-(CH_2-CF_2)-$, the Poly(vinylidene fluoride) (PVDF) - see Fig.5, and the family, e.g. P(VDF-TrFE). PVDF is exploited as a piezoelectric if poled, i.e. quenched under an applied voltage. The ferroelectricity is of a conformational origin, i.e. due to rotation of chains' segments from the nonpolar cis isomer to the polar trans one. The molecular units of PVDF have net dipole moments (Fig. 5, right) and the dipoles are aligned in parallel in the trans form, which gives the transverse dipole moment to the whole chain. Polarizability of PVDF comes from dipole moments of the ionic origin; it is due to the electronegative fluorine and the electropositive hydrogen.

Typical samples of PVDF are polymorphous, containing an amorphous material and one or more crystalline phases, one of which is the ferroelectric. Here, the chains possess the all-trans conformation (Figure 5, left), they are tightly packed in a quasi-hexagonal polar $C_{2V}$ structure. The chains can crystallize in parallel rows and, in the ferroelectric state, the dipoles of all chains are aligned along a twofold crystalline axis, resulting in a macroscopic polarization (Fig.5, right). There is no clear Curie-Weiss like phase transition, it seems to be hidden by the melting point at 170°C. Switching of the polarization is accomplished by applying a large electric field opposing to the remnant polarization.

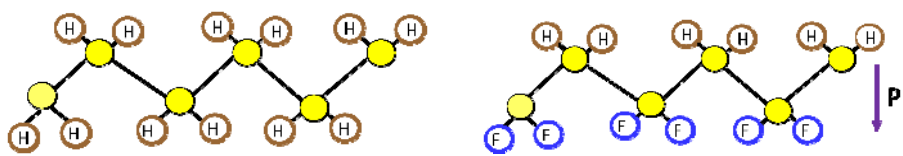

Figure 5. $(CH_2)_x$ chain, left; PVDF chain, right.

## 4.2. Future ferroelectric conjugated polymers.

The natural attempt to enhance the magnitude and the speed of the polarization is to turn to conjugated polymers, to exploit their fast π-electronic manifold instead of common ions. Then the ferroelectricity will be of electronic origin, i.e. appearing thanks to primary redistribution of π-electronic density rather than to displacement of ions. Beyond a set of principally new features, the responsive π- electronic systems can greatly enhance the dielectric permittivity ε - theoretically [12] by a factor of $(\omega_p/\Delta)^2 \sim 10^2$ ($\omega_p$ is the plasma frequency, $\Delta$ is the gap in the electronic spectrum).

Up to now, the activity in physics and applications of conjugated polymers was centered on their ability to absorb and emit the light in the optical range of about 2eV which is naturally provided by typical gaps and excitons' energies of π-electrons, hence the focus on conjugated polymers (see e.g. [35] and references therein). These materials are already close to applications and even commercialized in optical and electro-optical applications [36]. If one succeeds to use their fast π-electrons making the ferroelectric, it will be possible also to relate transient ferroelectric processes and the visible range optics, thus giving additional functionalities to these materials [27-28]. After our arguments, it is natural to search for the necessary materials within the general class of functionalized polyacetylens [37,38]. More specifically, they are called mono- or di-substituted polyacetylenes, depending on the number of hydrogen atoms substituted by some radicals - see reviews [39,40], or schematically as the $(AB)_x$ conjugated polymers.

In recent years, several laboratories over the world have been working with these perspective materials [37-40], obtaining the LED [41] and even lasing [40], but no attempts have been made to tests for the ferroelectricity, so our recent calls were in vain [27,28]. We admit that among these studies the ferroelectric polyacetylenes have been obtained [38], in the course of the beautiful program of hierarchical polymers [37]. But the polarization was very weak, being a consequence of polarity of ligands, similarly to ferroelectric liquid crystals. Here we are discussing an opposite strong effect coming from electrons of the backbone.

In early 80's, an existence of a conjugated analog of PVDF - $(CHCF)_x$ has been briefly mentioned, and we suggested [24,25] it as a proper $(AB)_x$. But that may not work for our goals: the spontaneous bond dimerization may not be generated because of a too strong effect of the substitution H→F. As we shall demonstrate in the Ch. 7, the ferroelectric state appears as a threshold effect. The spontaneous dimerization will not be generated if the built-in gap from the site difference exceeds the optimal Peierls gap in this material. The precaution is that the built-in gap from the site dimerization should be sufficiently small to leave the space for the contribution from the bonds dimerization gap. Thus, the only tiny difference of radicals in the di-substituted polyacetylenes shown in Fig.6 (left) will most probably fit the criterion. And indeed, the reliable indications on physics of solitons do confirm that [41]. But for two types of mono-substituted polyacetylenes shown in Fig.7 (center), a strong difference between the remnant hydrogen and the substitution may not (but not exclude to) satisfy the necessary conditions. Another requirement is the trans configuration. Thus the cis form shown (may be arbitrarily) in Fig.7, (lower panel, center) might be a pyroelectric: the forward and backward directions of polarization along the chain are not equivalent because the neighboring bonds are not the same *a priori*. The versatility of substituted polyacetylens will certainly allow to find the necessary conditions.

In absence of fine structural analysis, the presence of the spontaneous dimerization, hence of the polarization, is questionable. But we can firmly deduce that from identification of solitons: these kinks of the bonds alternation can exist only in presence of the spontaneous symmetry breaking. Based on the early theory of the "combined Peierls effect" [24,25] (described below in Ch. 7), we predicted an existence of solitons with non-integer variable charges, both with and without spin, which are the walls separating domains with opposite electric polarisation. (This notion became popular thanks to the nick-name of the (AB)x polymer coined in [26]). The physics of solitons will serve to relate transient ferroelectric processes and the visible range optics.

We also indicated the class of conducting polymers – the substituted polyacetylenes - where the ferroelectricity should be present [27,28]. One such a polymer has already been studied for nonlinear optical properties [41], but not yet tested for the ferroelectricity. Impressive set of combined ESR - optical experiments and IRAV (infra-red-active-vibrations), see Fig.6, indicate spin carrying solitons which also possess an electric charge; it is surprising from experience of the conventional polyacetylene but absolutely natural for the $(AB)_x$ case. The complete optical characterization of (still one) substituted polyacetylene [41,42] provides indirect but convincing proofs for spontaneous bonds dimerization via spectral signatures of solitons.

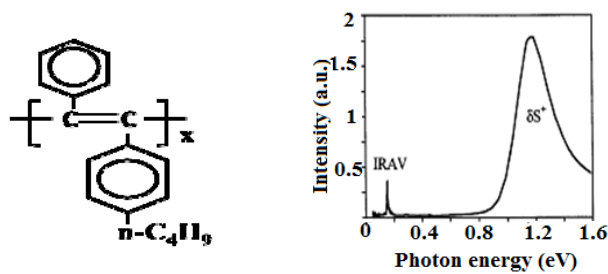

Figure 6. Left: the structure of the existing, and promising for the ferroelectricity, di-substituted polyacetylene. Right: the optical feature of solitons as an indication for presence of the spontaneous bond dimerization, after [41].

A vast variety of synthesis of substituted polyacetylenes has been already demonstrated [37-40]. Fig. 7 shows two among numerous existing compounds which can be at least the pyroelectrics. At the left panel the calculated structure and observed packing for one kind of mono-substituted polyacetylene (center, upper panel) is presented. Another modification of a mono-substituted polyacetylene (center lower panel) gives rise to a visible-light lasing effect (right panel) [42]. Notice the very sharp peak with the four times increase of the photo-luminescence intensity and a very narrow line width, less than 10 nm. Notice that the carbon backbone has a cis isomerization (probably just guessed) – the degeneracy of bonds' dimerization is already lifted, the polarization is weakly frozen. While this particular compound may not be ferroelectric (still, to be tested !) because of a strong difference between the hydrogen and the substitution radical (as we have discussed above) still the way is opened to the ferroelectrically tunable laser or LED.

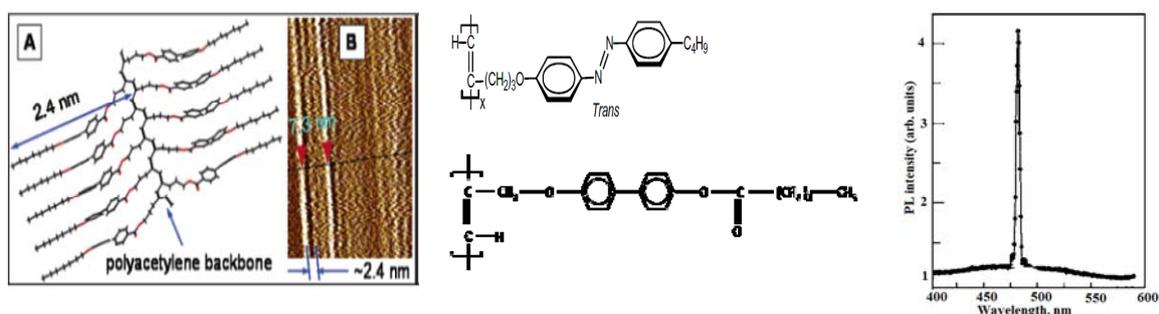

Figure 7. Left: Calculated chain structure (A) and the observed packing (B) of a mono-substituted polyacetylene (center, upper panel), taken from [39]. Right: Lasing effect at T=12 K in a mono-substituted polyacetylene (center, lower panel), the experimental points are taken from [42].

**4.3. Polar edges of graphene?**

We can speculate also on the ferroelectricity at zigzag edges of graphene nano-ribbons, see Fig.8. Here, the build-in dimerization of sites is originated by difference in surroundings of carbon atoms at the edge and in the bulk, while the spontaneous dimerization of bonds can result from the Peierls instability in electronic edge states. Effects of the size and the edge shape of graphene ribbon upon the electronic edge states were modeled numerically in [43]. It was found that peaks from the edge states in the total density of states show up clearly for a ribbon width below about 10 nm (comprising some 50 zig-zag carbon chains). The wave function of edge states are localized within a few chains which corroborates with calculations [44] for benzene ring based polymers. Hence, for a nano-ribbon with a width of about or more than 10 carbon rings, the edge states on the opposite sites of the ribbon will not interact and the ferroelectricity can be protected.

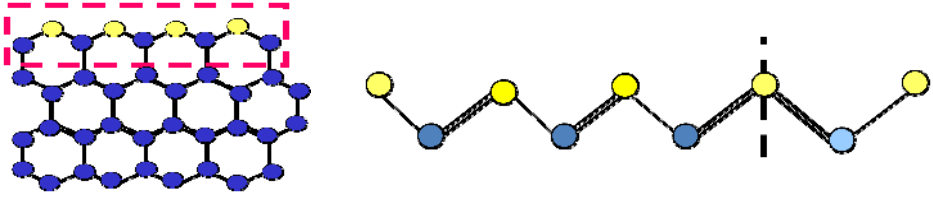

Figure 8. Analogy of edge states of graphene and (AB)x polymer.

The open question is: will the bond alternation appear at a reasonable temperature? The answer depends on the passivating radicals for the ribbon edges as well as on the coupling constant $\lambda$ of the edge state with lattice deformations. By now, theoretical and experimental estimations for $\lambda$ strongly diverge ranging from $\lambda \sim (3-5) \cdot 10^{-2}$ [45,46], to $\lambda \sim 0.5-1.5$ [47]. This question is still waiting for clarification.

## 5. Organic crystals.

The ferroelectricity in carbon based materials was first discovered [5] (while the name was not coined) in so called Bechgard-Fabre salt: quasi-1D organic conductors $(TMTTF)_2X$, where $X=AsF_6$, $SbF_6$, $PF_6$, $ReO_4$ and TMTTF is the molecule shown in Fig.10. Soon, the observation was expanded to layered compounds (see [30] and rfs. therein). In all cases, the ferroelectric subfamily had superconducting and magnetically ordered neighbors, while at much lower temperatures.

The part in permittivity contributed from the purely ionic displacements is expected to be $\varepsilon \sim 10^1 (T_c/|T-T_c|)$, which is at least three orders of magnitude below experimental value $\varepsilon \sim 2.5 \times 10^4 (T_c/|T-T_c|)$. Near the phase transition temperature (the highest is $T_{FE} \approx 150K$), the dielectric permittivity reaches a gigantic value $\varepsilon \sim 10^6$ [5,13] and even well below $T_{FE}$ it is as high as $\varepsilon \sim 10^3$, see Fig.9 - left. Notice the unusual for ferroelectrics genuine second order phase transition for all four materials as confirmed by the perfect Curie law, Fig. 9 (right). There were no attempts to measure the remnant polarization or to see the hysteresis loop which is not easy in this well conducting material.

The direct experimental proof of a ferroelectric state was presented by local measurements of the second harmonic generation; the results have been published [30] for a layered compound $\alpha$-$(ET)_2I_3$ and reported vaguely for $(TMTTF)_2X$ [48]. At presence of the frozen polarization, the inversion symmetry is lifted, then the polarization $P(E)$ can be written as:

$$P = \frac{\varepsilon}{4\pi}E + \chi_2 E^2 \qquad (5)$$

where the term with $E^2$ is allowed only if the inversion symmetry is lifted; that gives rise to the second harmonic generation in optics.

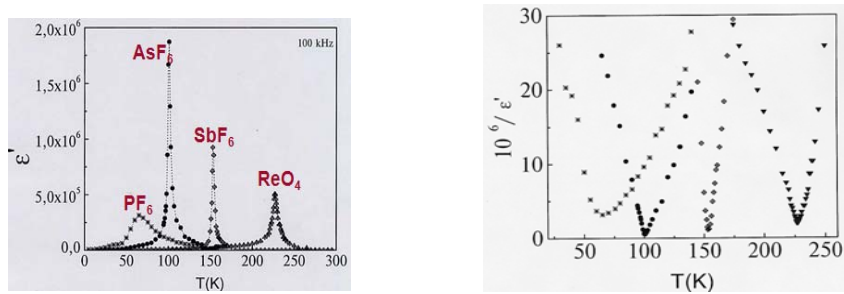

Fig.9. Real part of the permittivity $\varepsilon'$ and its inverse in $(TMTTF)_2X$, taken from [5].

The TMTTF molecules form the stacks intercalated by columns of counter-ions X which are placed against every second pair of molecules (see Fig.10, left). The impact of columns upon stacks provokes the small built-in dimerization of intermolecular bonds, while leaving the molecules to be equivalent. Below the transition, there is a spontaneous symmetry breaking visualized by displacements of counter ions (indicated by arrows at Fig.10, center and right) which follow and stabilize the charge disproportination between TMTTF molecules [49,50]. Contrary to traditional ionic ferroelectrics and even other organic ones, the new state keeps a noticeable conductivity and preserves free-spin paramagnetism, giving rise to a "ferroelectric narrow gap Mott semiconductor", see [12] for a review. Free carriers screen the surface polarization preventing formation of the domain structure, thus allowing for a rare mono-domain state and leading to the fast re-polarization.

As always, general principles apply only to on-chain effects leaving the 3D arrangement to depend on ill-predictable interchain interactions. The global state will be ferroelectric if on-site dimerizations of all chains are in phase; that can seen via the counter-ions X displacing in the same direction. The global state will be antiferroelectric, if on-site dimerizations of neighboring chains are out-of-phase; that is well seen via the counter-ions X in different columns displacing in opposite direction. Encouragingly, among the whole family (TMTTF)$_2$X only the case of X=SCN is antiferroelectric. While less interesting in applications and less spectacular in showing only a kink in the permittivity, this case is very fortunate for X-ray studies [49].

The microscopic picture of ferroelectricity in organic crystals [12] will be discussed below in Ch. 7.

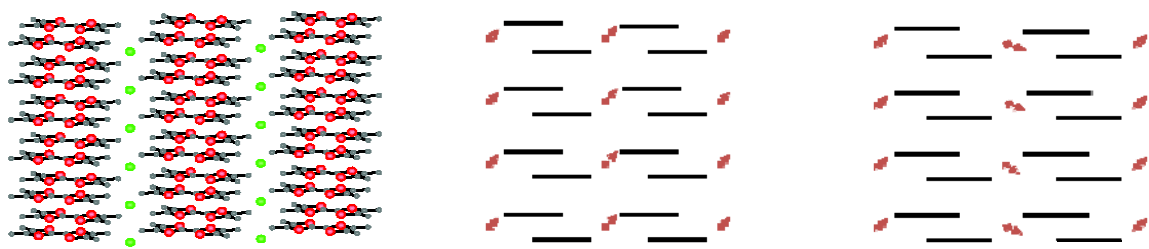

Fig.10. Structure of (TMTTF)$_2$X (left). Formation of spontaneous site dimerization for ferroelectric (center) and antiferroelectric (right) states accompanied by shifts of the X-ions toward one of the TMTTF molecule in the pair.

## 6. Ferroelectricity in bi-molecular donor-acceptor chains

### 6.1. Thermodynamic phase transition

Bi-molecular mixed-stack organic crystals like the TTF-CA consist of stacks of alternating donors D=TTF and acceptors A=CA as shown in Fig.11, left. The ferroelectricity appears as a result of the neutral-ionic transition [6,10,15,16] which brings to life, or at least sharply increases, both types of dimerizations, usually that comes simultaneously.

The site dimerization is present already above the phase transition at $T_{NI}$ because of different donor D and acceptor A molecules occupying the alternating sites, but it substantially increases by the further charge transfer ρ within the dimer (D$^{\rho}$-A$^{-\rho}$). In the quasi-neutral high temperature phase the charge transfer $\rho_N$ comes from the hybridization of D and A bands opposed by the Coulomb energy paid for the ionization; the molecules are weakly charged and equidistant. In the ionic phase below the transition temperature ($T_{NI}$=81 K for TTF-CA), an additional spontaneous charge transfer takes place to a higher level $\rho_I > \rho_N$.

The ferroelectricity below the neutral-ionic transition in TTF-CA was confirmed by the sharp peak anomaly in temperature dependence of the permittivity [51] and by well defined hysteresis loop in the polarization curve [10], as well as by X-ray diffraction studies [15]. The electronic component of the polarization happens to be 20 times higher than the ionic one. Moreover, the electronic and the ionic polarizations point in opposite directions.

For such a thermodynamic phase transition the primary effect is the redistribution of the charge density ρ with no symmetry breaking, hence an isomorphic transition which could be described by the single parameter $q=\rho-\rho_N$. Such a transition should be, and it is, of the 1$^{st}$ order; but the critical increase of the dielectric constant approaching $T_{NI}$ is observed [10,51], so the transition must be close to the critical point (by analogy to liquid-gas transitions).

Fortunately, the real situation in TTF-CA is more complicated since another degree of freedom enters the game: alternating molecular displacements *h* appear in the ionic phase, and this dimerization gives rise to a symmetry breaking. Namely, the charged molecules shift forming the alternating dimers which can be either Coulomb or spin-Peierls instabilities. As all inversion and mirror symmetries are lifted, this state must be ferroelectric in accordance with the general scheme outlined above in Ch.3.

Taken alone, this transition would be of the 2$^{nd}$ order but it is not the case since *h* appears together with the jump in $\rho$ from 0.3 to 0. 52.

The ground state degeneracy and the coexistence of neutral and ionic segments give rise to several types of microscopic domain walls – the solitons [22,23] which effects are actively discussed in the modern experimental literature. Because of equivalent symmetries, there is a direct correspondence among these solitons and those in organic conductors and conjugating polymers which we shall discuss in other sections.

Fig.11(right) illustrates three thermodynamic regimes of the free energy landscape depending on the temperature *T*. At $T>T_{c1}$, there is the stable N state at *q=0, h=0* and two shallow metastable I states at finite *q* and *h*. At $T=T_{c1}$ both N and I states have the same energy, separated by the barrier. At $T_{c2}<T<T_{c1}$ the I state is stable and the N state is metastable. At $T=T_{c2}$ the barrier disappears, so at $T \leq T_{c2}$ the N state becomes absolutely unstable.

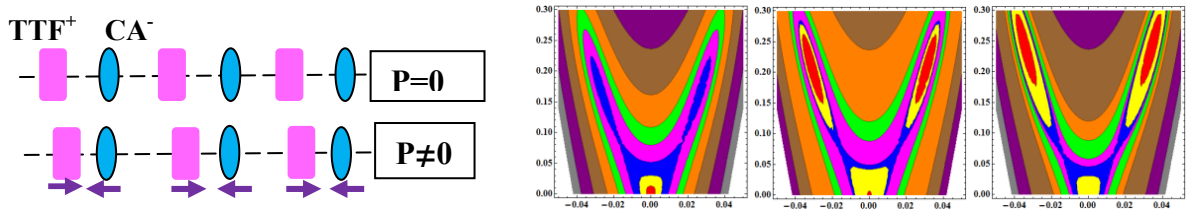

Figure 11. Left: Neutral-ionic transition (a-c) in bi-molecular donor-acceptor chain TTF-CA. Right: 2D contour plots for the phenomenological free energy *W(q,h,T)* as functions of *q* (the vertical axis) and *h* (the horizontal axis) at different T. Left: $T>T_{c1}$, center: $T= T_{c1}$, right: $T_{c1}>T>T_{c2}$. Similar pictures appear if the concentration of optically pumped excitons (see Ch.6.2) is monitored instead of the temperature.

## 6.2. Pump induced ferroelectric neutral-ionic transition.

One mainstream of the modern condense matter physics exploits transformations among cooperative electronic states performed by short optical pulses [53-56]. By these "pump induced phase transitions" (PIPT), various symmetry broken ground states are being tested: crystallization of electrons (charge order with ferroelectricity) or of electron-hole pairs (charge/spin density waves – CDW,SDW), the superconductivity, Peierls and Mott insulators. In most experiments on PIPTs, the electrons are excited to a high energy with their subsequent very fast cooling down to a quasi-equilibrium e-h population which provokes events of a subsequent evolution. A new situation can take place if the photons are tuned in resonance with excitations, whatever is their origin – intra-molecular excitons (IME) or bound electron-hole pairs – charge transfer excitons (CTE). This is the case of an experimentally elaborated example of optically provoked transformation between neutral and ionic phases in organic crystals like TTF-CA [17-20]. In view of a pivotal role of excitons in conjugated polymers (see [57] for a theory review), this technique deserves the attention in science and applications of conjugated polymers, so we devote here some space to this subject. The big differences in excitons' energies: 2.4eV for IME and 0.6 eV for the CTE reflects their different nature and calls for different approaches in both, theory and experiment.

Recently, we contributed to the field of PIPTs by noticing a possibility of macroscopic quantum coherence in ensembles of briskly pumped excitons [52]. With experimentally achieved high concentration of the excitons (up to 10%), the quasi-condensate of excitons should appear as a macroscopic quantum state which then evolves interacting with other degrees of freedom. The common case of pumping to unbound electrons and holes is not excluded from our scenario, provided the early cooling leads first to formation of excitons; this is what is known to take place in light emitting polymers [57] or in conventional semiconducting lasers operating at low temperature.

Particularly, we were interested in effects of excitons' self-trapping [58], akin to self-focusing in the nonlinear optics or to formation of polarons from electrons. The locally enhanced density of excitons can surpass a critical value to trigger a phase transformation in another coupled degree of freedom, even if the mean density *n* is below the required threshold. For PIPT in TTF-CA in case of the resonance pumping to IME, the phenomenological model and its numerical exploration have been presented in [52].

The pumping of excitons changes the energy landscape already for $T>T_{c1}$. There are three critical levels of the pumping density: two of them $n_{c1}$ and $n_{c2}$ are from thermodynamics. They correspond to the 1$^{st}$ and the 2$^{nd}$ order phase transitions, and the third one - $n_d$ comes from dynamics. At $n_{c1}$ the energies of both N and I states are equal and separated by the barrier (to be compared with $T=T_{c1}$). At $n=n_{c2}$, the barrier disappears, the N phase becomes absolutely unstable (to be compared with $T=T_{c2}$). At $n=n_d$, the N phase still exists as a metastable state, the barrier is here, but its height drops down to $W_b=0$ – the ground state level before the pumping. Then the barrier can be overcome by inertia in the course of large-amplitude oscillations after the short pulse of pumping. At very low pumping $n<<n_{c1}$, the concentration of excitons is not sufficient to initiate the development of *h*. At higher pumping, but still at $n<n_{c1}$ the self-trapping develops self-consistently in the excitons' wave function *ψ* and in *q*, resulting at later time in the appearance of *h*. Formation of a sharply localized long living domain for all three variables is due to self-trapping effects in spite that the pumping is still subcritical $n<n_{c1}$.

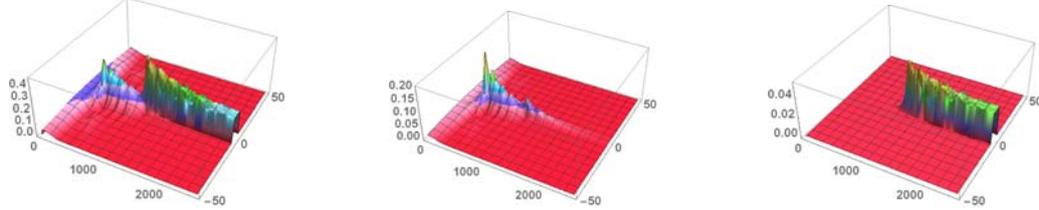

Figure 12. Charge transfer (left), density of excitons (center), spontaneous deformation (right) calculated for the femtosecond initial pumping. Notice that the spontaneous deformation appears much later than the charge transfer.

## 7. Elements of theory.

### 7.1. Landau theory of the ferroelectric phase transition

In general, a ferroelectric undergoes a thermodynamic phase transition upon cooling towards a lower symmetry phase, with the appearance of an order parameter with at least two equivalent energy states. This phase transition phenomenologically can be described by the free energy $W(P,E,T)$, with the polarization ***P*** as the order parameter [1]. Its simplest form is

$$W(\boldsymbol{P}, \boldsymbol{E}, T) = -\boldsymbol{P}\boldsymbol{E} + \frac{a}{2}(T-T_c)\boldsymbol{P}^2 + \frac{b}{4}\boldsymbol{P}^4 + \frac{c}{6}\boldsymbol{P}^6 + l^2(\boldsymbol{\nabla}\boldsymbol{P})^2 \quad (1)$$

Here ***E*** is the external electric field, *T* is the temperature. The 6$^{th}$ order term in *P* is kept because frequently the ferroelectric transition is of the 1$^{st}$ order corresponding to the negative $b<0$. Variation of W with respect to ***P*** results in the equation:

$$-\boldsymbol{E} + a(T-T_c)\boldsymbol{P} + b\boldsymbol{P}^3 + c\boldsymbol{P}^5 = 0 \quad (2)$$

For the ground state ***E***=0, ***P***=cnst, there are the following solutions of Eq.(2):
*1. Second order phase transition* (Fig.14, left): if $b>0$, the term ~$P^5$ does not bring new features and can be neglected. At $T>Tc$: $P_{PE}=0$ - the paraelectric phase; at $T<Tc$: $P_{FE} = \pm\sqrt{(T_c-T)a/b}$ - the ferroelectric phase. The dielectric susceptibility has a form

$$\varepsilon_{PE} = 4\pi \frac{dP}{dE} = \frac{4\pi k}{a(T-T_c)} \quad (3)$$

where k=1 for the paraelectric phase and k=1/2 for the ferroelectric phase.
For the case of 1st order phase transition (Fig.14, right) at $b<0$, the term $cP^5$ should be kept and the solutions yield: at $T>Tc$, $P_{PE}=0$ - the paraelectric phase; at $T_0<T<Tc$ - paraelectric and ferroelectric phases coexist; at $T<T_0$ - only the ferroelectric phase is stable. The polarization in ferroelectric phase in both temperature regions is

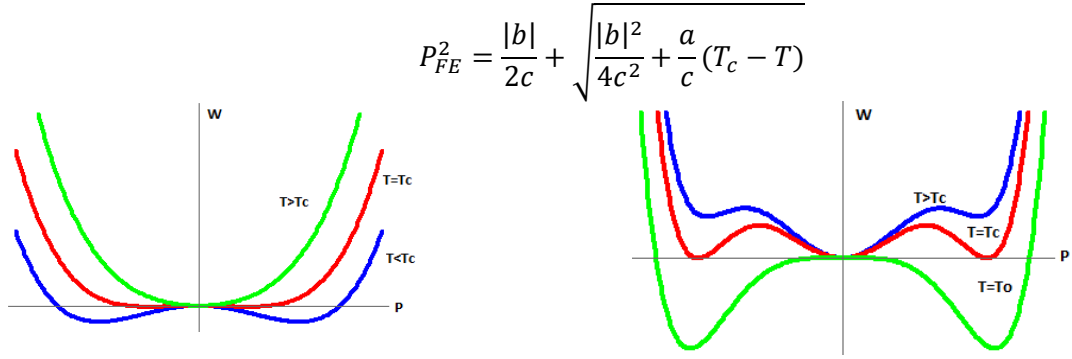

Fig.13. Free energy plots for the second (left) and first (right) order ferroelectric transition

### 7.2. Combined Mott-Hubbard state in organic crystals.

The microscopic picture of ferroelectricity in organic crystals is based on the spontaneous charge disproportination with forming of the favorable Mott-Hubbard state [5,11,12]. Bonds' dimerization doubles effectively the unit cell, thus changing the mean electronic occupation from ½ per molecule to 1 per dimer which opens the route to the Mott-Hubbard insulator [12,59,60]. A lesson is that this state is energetically favorable, so the system away from the Mott insulator mobilizes available degrees of freedom to fall into necessary conditions – here going from ¼ to ½ fillings.

The resulting combined Mott-Hubbard state can be viewed and described as a Wigner crystal of electrons – a periodical charge modulation which period incorporates just one electron. This picture is equivalent to a view of a so called *$4k_F$ charge density wave (CDW)* which is found in many organic conductors and is associated with the strong Coulomb repulsion, see [50] for rfs. At presence of any dimerization, the CDW profile *$A\cos(\pi/n+\varphi)$* becomes one-to-one commensurate with the lattice, but the lock-in phases for two types of dimerizations are sifted by *π/2*. Hence, the lock-in energies are *$W_b=U_b\sin\varphi$* and *$W_s=U_s\cos\varphi$* for dimerizations of bonds and sites correspondingly. The phase φ should be determined by minimization of the total free energy

$$W = W_{MH}(U) - U\cos(\varphi - \alpha) + \frac{K}{2}U_S^2 \, ; \; U = \sqrt{U_S^2 + U_B^2}, \, \tan\alpha = \frac{U_S}{U_B} \qquad (6)$$

where the term ~$U_s^2$ comes from the energy paid for lattice deformations and $W_{MH}(U)<0$ is the energy gain in the Mott-Hubbard state; it depends only on the total amplitude *U* and does not depend on the phase α. Notice that the spontaneous site dimerization appears as a threshold effect. The minimization of (6) gives the relation

$$U_B^2 + U_S^2 = \frac{1}{K} \qquad (7)$$

Hence, at *$U_B>K^{-1/2}$* no additional spontaneous site dimerization $U_S$ can be formed.

In the ferroelectric phase the ground state is still doubly degenerate between *φ=α* and *φ=α±π*. It allows for solitons which are the ±π phase kinks – 1D incarnations of "holons" and "doublons" with the charge ±*e*. These purely on-chain solitons should exist as conducting quasiparticles both above and below $T_{FE}$. In the ferroelectric phase with $U_S\neq 0$, this amplitude can change the sign Us↔-Us. Then the electronic system must also adjust its ground state from *α* to *–α*, i.e. between ferroelectric domains with opposite polarizations. The domain boundary requires for the phase soliton with the increment *δ = ±2α* which will concentrate the non-integer charge *q=±2(α/π)e* per chain.

These fractional solitons are nothing but elementary domain walls between on-chain (on-stack) domains of opposite ferroelectric polarizations. With the onset of the global 3D ordering, this type of solitons aggregate [25] to form the macroscopic domain walls between regions with opposite polarizations.

### 7.3. Combined Peierls effect in conducting polymers.

The simplest conducting polymer is the polyacetylene - $(CH)_x$ [61] where the high conductivity has been discovered by A. Heeger, A. McDiarmid, Y.W. Park and H. Shirakawa. Here, three among four carbon valence electrons form the σ bonds of the chain skeleton, leaving the fourth π electron free. With one electron per unit cell that should be a metal. But in a one-dimensional metallic chain, even a small electron-lattice interaction changes drastically the electronic properties. The basis for this understanding is the old Peierls effect which is known also as the SSH model. It states that a one-dimensional equally spaced chain with one electron per atom is unstable with respect to dimerization: the alternation of intermolecular distances. Spontaneus symmetry breaking results in the dielectric state where the elementary cell contains two atoms, hence there are two electrons per unit cell, the band becomes completely filled and the material becomes insulating. The gap $2\Delta_0 \sim D\exp(-1/\lambda)$ is opened at the Fermi-surface. Here $D$ is bare width of the electronic band and $\lambda$ is the electron-phonon coupling constant.

Even the bond-dimerized $(CH)_x$ chain cannot be ferroelectric which is prevented by persistence of the glide plane symmetry (1/2-period translation plus the mirror reflection) [61,62]. To get the ferroelectric, we need to invoke also the site dimerization, see Ch.3, which removes the glide plane symmetry opening the way for the on-chain polarization. Now, the material is insulating from the very beginning because of the ubiquitous site dimerization yielding the gap $\Delta_e$.

The critical condition for the ferroelectricity in an $(AB)_x$ chain is to acquire a spontaneous bond dimerization which opportunity is not obvious. Under the joint effect of extrinsic (built-in) $\Delta_e$ and intrinsic (spontaneous) $\Delta_i$ contributions, the total gap $\Delta_C$ in the electronic spectrum E(p) becomes [24]

$$E(p) = \sqrt{v_F^2 p^2 + \Delta_C^2} \qquad \Delta_c = \sqrt{\Delta_e^2 + \Delta_i^2} = min\{\Delta_0, \Delta_e\} \qquad (11)$$

Here $\Delta_0$ is the optimal Peierls gap for the parent, with equivalent sites, system. We see from (11) that the appearance of the spontaneous dimerization is the threshold effect: $\Delta_i$ will not be generated, if $\Delta_e$ already exceeds the wanted optimal Peierls gap $\Delta_0$. This result gives us immediately the chemistry precaution: to keep small the difference of ligands R and R′. The same holds for graphene nano-ribbons concerning the passivativing groups of their border, see Ch. 4.3, or atomic potentials for A and B atoms in an arbitrary $(AB)_x$ chain should not be very different.

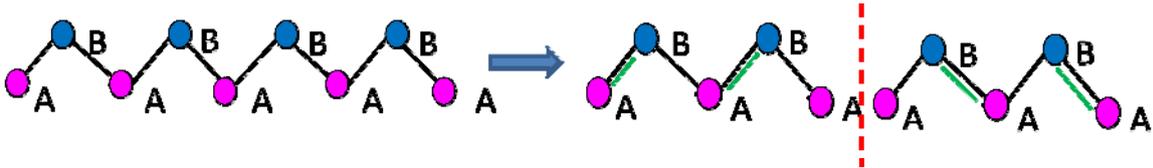

Fig.14. Appearance of spontaneous bond dimerization in $(AB)_x$ polymer. Left: dielectric state without spontaneous dimerization. Right: Ferroelectric state where both built-in and spontaneous dimerizations are present. Notice the doubly degenerated ground state.

Both the ground state and the stationary excited states like solitons and polarons can be obtained as minima of the energy functional W dependent on the function $\Delta_i(x)$ at $\Delta_e$=cnst [24,25]:

$$W = Tr \begin{vmatrix} -i\hbar v_F \partial_x & \Delta_e + i\Delta_i \\ \Delta_e - i\Delta_i & i\hbar v_F \partial_x \end{vmatrix} + \frac{K}{2}|\Delta_i|^2 \qquad (12)$$

We can easily arrive at the above result for the ground state and, after some math, to the solution for a soliton:

$$\Delta_i(x) = \sqrt{\Delta_0^2 - \Delta_e^2} \tanh\frac{x}{\xi_0} \; ; \; \xi_0 = \frac{\hbar v_F}{\Delta_0} \qquad (13)$$

On the plane $(\Delta_i, \Delta_e)$, the soliton trajectory is shifted from the diameter to the horde by the distance $\Delta_e$ (Fig.15, right). At presence of the soliton, the localized level is split off inside the gap, but contrary to the standard $(CH)_x$ case, the corresponding intragap states now are shifted from the gap center to $\pm\Delta_e$ - the sign depends on the level filling, see Fig.15, left and center. Moreover, now both the earlier charged spinless soliton with s=0 and the earlier neutral soliton with s=1/2 have non-integer charges $Q_{Ch}$ and $Q_S$ [24,25] which are determined by the chiral angle $\theta$:

$$\theta = \tan^{-1}\frac{\Delta_i}{\Delta_e}, \quad Q_{Ch} = 2e\frac{\theta}{\pi}, \quad Q_S = 2e\frac{\pi-\theta}{\pi} \tag{14}$$

At $\Delta_e=0$ we have $\theta=\pi/2$ and we return to standard result for the $(CH)_x$ chain with $Q_{Ch}=e$, $Q_S=0$.

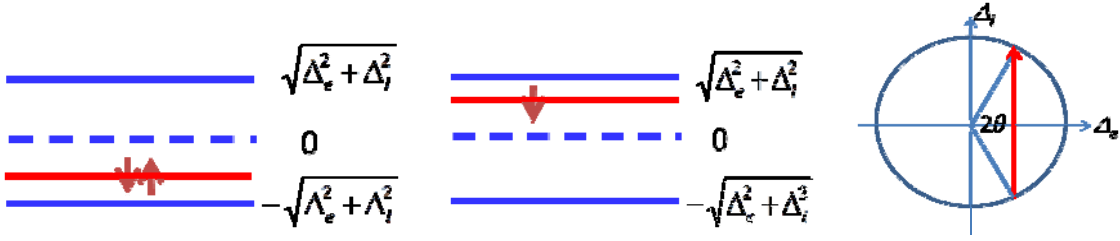

Figure 15. Left and middle: energy levels for a singlet (left) and spin 1/2 (center) solitons. Right: solitons trajectory on the plane ($\Delta_i,\Delta_e$); the ring corresponds to $\Delta_0 = \sqrt{\Delta_e^2 + \Delta_i^2}$.

### 7.3. Pump-induced phase transition in bi-molecular chain TTF-CA.

At the microscopic level theory of pump induced phase transition (PIPT) ([Yonemitsu], [Ishihara]) faces great difficulties, but over longer time scales the evolution is governed by collective variables and the phenomenological model can be developed. In this chapter we present a model for PIPT for pumping into intramolecular exciton, following [Bra-Kir_ECRYS]. Exciton's interaction with the inter-molecular charge transfer $\rho$ provoke a symmetry breaking lattice deformation $h$ resulting in the ferroelectric polarization. The double shape for the free energy (see Fig. 13) appears now from interactions of $\rho$ and $h$. The free energy is a functional of $q(x,t)=\rho-\rho_n$, ($\rho_n$ is the charge transfer in neutral state), lattice deformations $h(x,t)$ and the excitons' common wave function $\psi(x,t)$:

$$W(q,h,q_c,\psi) = \frac{a}{2}q^2 + \frac{b}{3}q^3 + \frac{c}{q_c}(q_c-q)h^2 + \frac{f}{2}h^4 + \frac{A}{2d^2}\left(\frac{\partial h}{\partial x}\right)^2$$
$$-gq|\psi|^2 + \frac{k}{2}|\psi|^4 + \frac{\hbar^2}{2md^2}\left|\frac{\partial \psi}{\partial x}\right|^2 \tag{15}$$

In (15) $q_c$ is the critical value of $q$ for reaching the instability in $h$. The energy minimum moves from $h=0$ at $q<q_c$ to $h=\pm h_{eq}$ at $q>q_c$; $g$ is the coupling constant of excitons with the charge transfer $q$, and $k$ is the repulsion energy of excitons. Terms with $x$ derivatives describe inhomogeneous states, $m$ is the exciton's kinetic mass. All lengths ($x,L,|\psi|^2$) are measured in units of the TTF-CA dimer size $d$ which is the lattice period in the N phase. Physical parameters and estimations can be found in [Bra-Kir_ECRYS].

$\psi(x,0)$ is normalized to the concentration per site $n$ of pumped excitons:

$$\int_{-L}^{L} dx\, |\psi(x,0)|^2 = nL \tag{16}$$

At a given homogeneous pumping $|\psi|^2=n=cnst$.
Variation of (15) gives us the equations for spacio-temporal evolution of the system:

$$i\hbar\frac{\partial \psi}{\partial t} + i\frac{\hbar}{\tau}\psi = -gq\psi + k|\psi|^2\psi - \frac{\hbar^2}{2m}\frac{\partial^2 \psi}{\partial x^2} \tag{17}$$
$$aq + bq^2 - ch^2 - g|\psi|^2 = 0 \tag{18}$$
$$\frac{1}{2\omega^2}\frac{\partial^2 h}{\partial t^2} + \frac{\gamma}{2\omega}\frac{\partial h}{\partial t} = \frac{s^2}{2d^2\omega^2}\frac{\partial^2 h}{\partial x^2} - \frac{f}{c}h^3 - h(q_c-q) \tag{19}$$

The pumping intensity is introduced via initial conditions:

$$q(x,0)=0, \quad h(x,0)=0, \quad \psi(x,0)=\sqrt{n}\cos\frac{\pi x}{2L}$$

## 8. Conclusions, perspectives and the road map.

### 8.1. Outlook for electronic ferroelectrics.

The microscopic picture of all electronic ferroelectric materials is based on coexisting symmetry lowering effects resulting in formation of a combined state characterized by two order parameters. If the dominant interactions are the electron-electron ones as in conducting organic crystals, it is the combined Mott-Hubbard state. In case of dominant electron-phonon interactions as in conducting polymers, it is the combined Peierls state. It appears that the formation of the ferroelectric ground state is a threshold effect with respect to the magnitude of the sites' dimerization.

Ferroelectric polarization has at least two stable states, being reversibly switched from one to another by application of an electric field. The regions with different orientations of the polarization vector may coexist within a ferroelectric sample, separated by ferroelectric domain walls; their microscopic nucleuses (on-chain domain walls) are the solitons, curiously with non-integer charges. They are on-chain conducting particles only above $T_c$; below $T_c$, they aggregate into macroscopic ferroelectric domain walls. Here, they do not conduct any more, but determine the ferroelectric depolarization dynamics. Charge carriers screen the electric field at the sample boundaries, eliminating the usual separation into domains of opposite polarization. In these unusual for common ferroelectrics circumstances, there are almost no hysteresis, easy repolarization and the astonishingly high permittivity already detected for organic crystals and preview for conducting polymers.

A special experimental advantage follows that the ac electric field alternates polarization by moving charged solitons. Through solitons' spectral features in visible and IR ranges, it opens a special tool of electro-optical interference. In return, the physics of these exotic solitons will serve to identify transient processes in ferroelectric polymers.

### 8.2. Road map to electronic ferroelectricity in conjugated polymers

Our proposal for the ferroelectricity is built upon symmetry analogies among representatives of three classes of materials: particular organic conductors and other organic materials with the neutral-ionic transitions which are already showing the "electronic ferroelectricity", and polymers from the family of the substituted polyacetylene with very broad synthetic options. We emphasized that such polymers already exist, and even studied for other purposes. Polymers of this type have been tried very successfully for the light emitting, but never tested for the ferroelectricity. There is a probability that already the first tried will show the effect, but actually a full scale project is necessary: the first materials can have a low transition temperature; they can show an antiferroelectric behavior, etc.

*A few words about the strategy.* There are several ways to detect the ferroelectricity, and each may be initially difficult, e.g. for noncrystalline samples or in presence of remnant conductivity. There is a lot of knowhow accumulated recently thanks to organic stack compounds. The ferroelectricity in a substituted polyacetylene is an ultimate consequence of the spontaneous Peierls dimerization known in the conventional polyacetylene. Our advice would be to concentrate on the proof of this bonds dimerization. We have argued for their existence indirectly through signatures of solitons. But there should be direct access by means of the X-ray scattering as it has been done for the generic polyacetylene in spite of difficulties from films disorientation [63-65]. Raman experiments can also give their word indirectly. The very welcomed microscopy of the second harmonics generation will help to register a local polarization even if it is averaged to zero in the bulk.

Traditionally, the ferroelectricity is tested by electrostatic devices: put the sample in a kind of a condenser and measure the polarization P and the permittivity ε. That may not work if a material is conducting (here, because of doping or of ions in the solvent) but then another unconventional method appears from [5,13]; this is how the ferroelectricity in $(TMTTF)_2X$ has been discovered. There, the kHz-MHz *ac* current is passed and ε is extracted from the impedance. Notice that junctions' problems have been reported. More traditionally, ε is obtained from the GHz techniques (the cavity method). This is how the "structuralless transitions" - to become later the ferroelectric ones - have been discovered (see the review [12] for the history). The anomaly in ε is weak at these high frequencies, so the effect may be registered only if there is a phase transition within the observable range of temperature. Finally, there may be approaches specific to polymers, such as poling of solutions:

orientation or crystallization under an applied electric field. Low temperatures will be necessary for the beginning: the effect must have a transition temperature. Cases of low temperature phases should not be overlooked.

The predictions face us with a responsibility to which the authors, the theorists, can only partly respond. That may explain why the earlier calls [27-28] have been left without any attentions. Most important will be a real time collaboration among chemists and experimentalists from condensed matter; it needs to be flexible because the objects will be quite unusual for specialists in traditional ferroelectricity, and feedback requests will be returned to the chemistry.

In conclusion, π-conjugated systems can support the electronic ferroelectricity. Effect is registered and interpreted in two families of organic crystalline conductors (quasi 1D and quasi 2D) and in mixed-stacks compounds. Mechanism is well understood as combined collective effects of Mott or (spin-) Peierls types. The energy is gained from the gap increase by the Mott, Peierls or spin-Peierls effects. Solitons will serve duties of re-polarization walls. An example of a must-be-ferroelectric polyene is already at hands. The design is symmetrically defined and can be previewed.

**Acknoweledgments:** The authors acknowledge the support from the grant NUST MISiS K3-2015-055**.**